\documentclass{cimento}

\usepackage{amsfonts,amsmath,amssymb,bm,graphicx,cite,makeidx,multicol,color}

\title{New bounds on neutrino magnetic moment
and re-examination of plasma effect in neutrino spin light}

\author{A.~V. Grigoriev\from{ins:SINP},
A.~V. Lokhov\from{ins:MSU}, A.~I. Studenikin\from{ins:JINR}\thanks{studenik@srd.sinp.msu.ru} \atque A.~I. Ternov\from{ins:MIPT}}

\instlist{ \inst{ins:SINP} Skobeltsyn Institute of Nuclear Physics, Moscow State University, 119991 Moscow, Russia \inst{ins:MSU}
Department of Theoretical Physics, Moscow State University, 119991 Moscow,  Russia \inst{ins:JINR} Department of Theoretical Physics,
Moscow State University, 119991 Moscow, Russia and \\Joint Institute for Nuclear Research, 141980 Dubna, Russia \inst{ins:MIPT} Department
of Theoretical Physics, Moscow Institute for Physics and Technology, 141700 Dolgoprudny, Russia}

\PACSes{\PACSit{13.15.+g} {Neutrinos interactions}
        \PACSit{95.30.Cq} {Elementary particles in astrophysics} }
\shorttitle{New bounds on neutrino magnetic moment and plasma
influence on spin light}
\begin{document}

\maketitle

\begin{abstract}
Recent discussion on the possibility to obtain more stringent bounds on neutrino magnetic moment has stimulated new interest to possible
effects induced by neutrino magnetic moment. In particular, in this note after a short review on neutrino magnetic moment we re-examine the
effect of plasmon mass on neutrino spin light radiation in dense matter. We track the entry of the plasmon mass quantity in process
characteristics and found out that the most substantial role it plays is the formation of the process threshold. It is shown that far from
this point the plasmon mass can be omitted in all the corresponding physical quantities and one can rely on the results of massless photon
spin light radiation theory in matter.
\end{abstract}

\section{Neutrino magnetic moment}
Neutrino magnetic moments are no doubt among the most well
theoretically understood and experimentally studied neutrino
electromagnetic properties. \cite{GiuStu09,Stu09_GiuStu10}

As it was shown long ago \cite{MarSanLeeShrFuj77_80}, in a wide
set of theoretical frameworks neutrino magnetic moment is
proportional to the neutrino mass and in general very small. For
instance, for the minimally extended Standard Model the Dirac
neutrino magnetic moment is given by \cite{MarSanLeeShrFuj77_80}:
\begin{equation}
    \mu_{ii}=\frac{3eG_F}{8\sqrt{2}\pi^2}m_i\approx
3.2\cdot{10^{-19}}(\frac{m_i}{1 eV})\mu_B.
     \label{magn_mom}
\end{equation}
At the same time, the magnetic moment of hypothetical heavy
neutrino (with mass $m_e\ll{m_W}\ll{m_{\nu}}$) is
$\mu_{\nu}=\frac{eG_{F}m_{\nu}}{8\sqrt{2}\pi^2}$ \cite{DvoStu}. It
should be noted here that much larger values for the neutrino
magnetic moments are possible in various extensions of the
Standard Model (see, for instance, in \cite{GiuStu09})

Constraints on the neutrino magnetic moment can be obtained in
$\nu-e$ scattering experiments from the observed lack of
distortions of the recoil electron energy spectra. Recent reactor
experiments provides us with the following upper bounds on the
neutrino magnetic moment:
$\mu_{\nu}\leq{9.0\times{10^{-11}}}\mu_B$ (MUNU collaboration
\cite{MUNUCollab05}), $\mu_{\nu}\leq{7.4\times{10^{-11}}}\mu_B$
(TEXONO collaboration \cite{TEXONO07}). The GEMMA collaboration
has obtain the world best limit
$\mu_{\nu}\leq{3.2\times{10^{-11}}}\mu_B$ \cite{Beda09}. Another
kind of neutrino experiment Borexino (solar neutrino scattering)
has obtained rather strong bound:
$\mu_{\nu}\leq{5.4\times{10^{-11}}}\mu_B$ \cite{Borexino08}. The
best astrophysical constraint on the neutrino magnetic moment has
been obtained from observation of the red giants cooling
$\mu_{\nu}\leq{3\times{10^{-12}}}\mu_B$ \cite{Raf90}.

As it was pointed out above the most stringent terrestrial
constraints on a neutrino effective magnetic moments have been
obtained in (anti)neutrino-electron scattering experiments and the
work to attain further improvements of the limits is in process.
In particular, it is expected that the new bound on the level of
$\mu_{\nu}\sim 1.5\times 10^{-11}\mu_B$ can be reached by the
GEMMA Collaboration in a new series of measurements at the Kalinin
Nuclear Power Plant with much closer displacements of the detector
to the reactor that can significantly enhanced the neutrino
flux(see \cite{Beda09}).

An attempt to reasonably improve the experimental bound on a neutrino magnetic moment was undertaken in \cite{WonLiLinPRL10} where it was
claimed that the account for the electron binding effect in atom can significantly increase the electromagnetic contribution to the
differential cross section in respect to the case when the free electron approximation is used in calculations of the cross section.

However, as it was shown in a series of papers
\cite{VolPRL10,KouStuPLB11,KouStuVolJETPLett11} the neutrino
reactor experiments on measurements of neutrino magnetic moment
are not sensitive to the electron binding effect, so that the free
electron approximation can be used for them.

\section{Magnetic moment and neutrino propagation in matter}
One may expect that neutrino electromagnetic properties can be
much easier visualized when neutrino is propagating in external
magnetic fields and dense matter. Also, neutrino propagation in
matter is a rather longstanding research field nevertheless still
having advances and obtaining a lot of interesting predictions for
various phenomena.

The convenient and elegant way for description of neutrino
interaction processes in matter has been recently offered in a
series of papers \cite{StuTerPLB05,StuJPA_06_08}. The developed
method is based on the use of solutions of the modified Dirac
equation for neutrino in matter in Feynman diagrams. The method
was developed before for studies of different processes in quantum
electrodynamics and was called as "the method of exact solutions"
\cite{SokTerRitNik} The gain from the introduction of the method
was sustained by prediction and detailed quantum description of
the new phenomenon of the spin light of neutrino in matter (the
$SL\nu$), first predicted in \cite{LobStuPLB03} within the
quasi-classical treatment of neutrino spin evolution. The essence
of the $SL\nu$ is the electromagnetic radiation in neutrino
transition between two different helicity states in matter.

The simplification of the process framework, such as use of the uniform, unpolarized and non-moving matter, neglect of the matter influence
on the radiated photon, makes the estimate of real process relevance in astrophysical settings far from the practical scope. In this short
paper we should like to make a step towards the completeness of the physical picture and to consider the incomprehensible at first glance
question of the plasmon mass influence on the $SL\nu$. The importance of plasma effects for the $SL\nu$ in matter was first pointed out in
\cite{StuTerPLB05}. The investigations already carried out in this area \cite{KuznMikh07} indicated that the plasmon emitted in the $SL\nu$
has a considerable mass that can affect the physics of the process. \textcolor{black}{However the calculation method used there does not
lead to the direct confrontation of the results \cite{KuznMikh07} with analogous for the $SL\nu$ \cite{StuTerPLB05}.}

To see how the plasmon mass enters the $SL\nu$ quantities we appeal to the method of exact solutions and carry out all the computations
relevant to the $SL\nu$. In this respect, in order to have the conformity we also set all the conditions for the task the same as for
corresponding studies on the $SL\nu$. In particular, we consider only the Standard Model neutrino interactions and take matter composed of
electrons.

In the exact solutions method, one starts with the modified Dirac
equation for the neutrino in matter in order to have initial and
final neutrino states, which would enter the process amplitude.
The equation reads as follows \cite{StuTerPLB05}:
\begin{equation}
    \{i\gamma_{\mu}\partial^{\mu}-\frac{1}{2}\gamma_{\mu}(1+\gamma^{5})f^{\mu}-m\}\Psi(x)=0, \label{eq:dirac}
\end{equation}
where in the case of neutrino motion through the non-moving and
unpolarized matter $f^{\mu}=G_{f}/\sqrt{2} \ (n,\textbf{0})$ with
$n$ being matter (electrons) number density. Under this conditions
the equation (\ref{eq:dirac}) has plane-wave solution determined
by 4-momentum $p$ and quantum numbers of helicity $s=\pm 1$ and
sign of energy $\varepsilon=\pm 1$. For the details of equation
solving and exact form of the wave functions
$\Psi_{\varepsilon,p,s}(\textbf{r},t)$ the reader is referred to
\cite{StuTerPLB05} and \cite{StuJPA_06_08}, here we cite only the
expression for the neutrino energy spectrum:
\begin{equation}
    E=\varepsilon\sqrt{(p-s\tilde{n})^{2}+m_{\nu}^{2}}+\tilde{n}, \ \ \tilde{n} = \frac{1}{2\sqrt{2}}G_F n.
    \label{eq:dispersion}
\end{equation}

The S-matrix of the process involves the usual dipole
electromagnetic vertex ${\bf \Gamma}=i\omega\big\{\big[{\bf
\Sigma} \times {\bm \varkappa}\big]+i\gamma^{5}{\bf \Sigma}\big\}$
and for given spinors for the initial and final neutrino states
$u_{i,f}$ can be written as
\begin{equation}
    S_{fi}=-(2\pi)^4\mu\sqrt{\frac{\pi}{2\omega L^3}}\delta(E_2-E_1+\omega) \delta^{3}({\bf p}_2-{\bf p}_1+{\bf
    k}) \overline{u}_{f}({{\bf e}},{\bf \Gamma}_{fi})u_i. \label{eq:amplitude}
\end{equation}
Here ${\bf e}$ is the photon polarization vector, $\mu$ is the
transitional magnetic moment and $L$ is the normalization length.
The delta-functions before spinors convolution part lead to the
conservation laws
\begin{equation}
    E_1=E_2+\omega; \ \   \bf{p_1}=\bf{p_2}+\bf{k}, \label{eq:conservation}
\end{equation}
with energies for the initial and final neutrinos $E_{1,2}$ taken
in accordance to (\ref{eq:dispersion}). For the photon dispersion,
for the purpose of our study it is sufficient to use the simplest
expression
\begin{equation}\label{photon dispersion}
    \omega=\sqrt{k^2+m^2_{\gamma}}.
\end{equation}

As it was discussed in our previous studies on the $SL\nu$
\cite{StuTerPLB05,StuJPA_06_08} the most appropriate conditions
for the radiation to manifest its properties are met in dense
astrophysical objects. This is the setting we will use further for
the process and in the case of cold plasma the plasmon mass should
be taken as
\begin{equation}\label{plasmon mass}
    m_{\gamma}=\sqrt{2\alpha}(3\sqrt{\pi}n)^{1/3}.
\end{equation}
The numerical evaluation at typical density gives
$m_\gamma\sim{10^8 eV}$, while the density parameter
$\tilde{n}\sim{10^4 eV}$.

\section{Plasmon mass influence}
Let us now consider the influence of dense plasma on the process
of spin light of neutrino. Similarly to the original spin light
calculation we consider the case of initial neutrino possessing
the helicity quantum number $s_1=-1$ and the corresponding final
neutrino helicity is $s_2=1$. Using the neutrino energies
(\ref{eq:dispersion}) with corresponding helicities one can
resolve the equations (\ref{eq:conservation}) in relation to
plasmon momentum which is not equal to its energy since we take
into account the dispersion of the emitted photon in plasma
(\ref{photon dispersion}).

For convenience of calculations it is possible to use the following simplification. In most cases the neutrino mass appeared to be the
smallest parameter in the considered problem and it is several orders smaller then any other parameter in the system. So we could first
examine our process in approximation of zero neutrino mass, though we should not forget that only neutrino with non-zero mass could
naturally possess the magnetic moment. This our simplification should be considered only as a technical one. It should be pointed here that
in order to obtain the consistent description of the $SL\nu$ one should account for the effects of the neutrino mass in the dispersion
relation and the neutrino wave functions.

From the energy-momentum conservation it follows \cite{KuznMikh07}
that the process is kinematically possible only under the
condition (taking account of the above-mentioned simplification):
\begin{equation}\label{threshold}
    \tilde{n}p>\frac{{m^2_\gamma}}{4}.
\end{equation}

Provided with the plasmon momentum we proceed with calculation of
the $SL\nu$ radiation rate and total power. The exact calculation
of total rate is an intricate problem and the final expression is
too large to be presented here. However one can consider the most
notable ranges of parameters to investigate some peculiarities of
the rate behavior.

First of all we calculate the rate for the case of the $SL\nu$
without plasma influence. This can be done by choosing the limit
$m_{\gamma}\rightarrow 0$ and the obtained result is in full
agreement with \cite{StuTerPLB05}:
\begin{equation}\label{gammaSL}
    \Gamma=4{\mu}^2\tilde{n}^2(\tilde{n}+p).
\end{equation}
From (\ref{gammaSL}) one easily derives the $SL\nu$ rate for two
important cases, \emph{i.e.} high and ultra-high densities of
matter just by choosing correspondingly $p$ or $\tilde{n}$ as the
leading parameter in the brackets. While neutrino mass is the
smallest quantity, our system fall within the range of
relativistic initial neutrino energies.

The corresponding expression for the total power also covers high
and ultra-high density cases \cite{StuTerPLB05} as well as the
intermediate area where the density parameter and the neutrino
momentum are comparable:
\begin{equation}\label{IntensSL}
    I=\frac{4}{3}{\mu}^2\tilde{n}^2(3\tilde{n}^2+4p\tilde{n}+p^2).
\end{equation}

If we account for the plasma influence (thus, $m_\gamma \neq 0$)
on the $SL\nu$ we can discuss two important situations. One is the
area of parameters near the threshold, and the other is connected
with direct contribution of $m_\gamma$ into the radiation rate
expression. The later case is particularly important for this
study, because it fulfill the aim of the present research in
finding the conditions under which the plasmon mass can not be
neglected.

For physically reliable conditions the density parameter usually
appears to be less then the plasmon mass, which in its turn is
less then the neutrino momentum: $\tilde{n}\ll{m_\gamma}\ll{p}$.
Obviously the threshold condition (\ref{threshold}) should be
satisfied. As we consider the conditions similar to different
astrophysical objects it is natural to use high-energy neutrino.

Using the series expansion of the total rate one could obtain the
rate of the process in the following form:
\begin{equation}\label{gammaSLseries}
    \Gamma=4{\mu}^2{p} \tilde{n}^2(1+6\lambda+4\lambda\ln\lambda),
\end{equation}
where $\lambda=\frac{m^2_\gamma}{4\tilde{n}p}<{1}$. Approaching
the threshold ($\lambda\rightarrow{1}$), the expansion
(\ref{gammaSLseries}) becomes inapplicable, however it is correct
in rather wide range of parameters with $m_\gamma\ll{p}$ and
$\tilde{n}\ll{p}$. Near the threshold the the total rate can be
presented in the form $\Gamma\sim{(1-\lambda)}$ but the exact
coefficient is too unwieldy to be presented here.

Concerning the power of the $SL\nu$ with plasmon, one can use the
expansion:
\begin{equation}\label{intensSLseries}
    I=\frac{4}{3}{\mu}^2{p}^2 \tilde{n}^2(1-6\lambda-57\lambda
    \frac{\tilde{n}}{p}-12\lambda\frac{\tilde{n}}{p}\ln\lambda).
\end{equation}
The expression (\ref{intensSLseries}) is correct only if the
system meets the requirement $\lambda\ll{1}$. Otherwise one should
use higher orders of quantity $\frac{m^2_\gamma}{p}$ in the
expansion to achieve a reliable value of intensity. Near the
threshold the power has the same dependence on the "distance" from
the threshold $(1-\lambda)$ as the rate of the process.

\section{Conclusion}
There is an increasing interest to neutrino electromagnetic
properties and neutrino magnetic moments in particular. This
interest is stimulated, first by the progress in experimental
bounds on magnetic moments which have been recently achieved, as
well as theoretical predictions of new processes emerging due to
neutrino magnetic moment, such as the $SL\nu$ and a believe in its
importance for possible astrophysical applications.

Further developing the theory of the spin light of neutrino, we
have explicitly shown that the influence of plasmon mass becomes
significant (see (\ref{gammaSLseries}) and (\ref{intensSLseries}))
when the parameter $\lambda$ is comparable with $1$, this
corresponds to the system near the threshold. As soon as the
quantity $\lambda\ll{1}$ (so the system is far from the threshold)
one can use either $SL\nu$ radiation rate and total power from
\cite{StuTerPLB05} or their rather compact generalizations
(\ref{gammaSLseries}) and (\ref{intensSLseries}) where the plasmon
mass is accounted for as a minor adjustment.

Since high energy neutrinos propagating in matter could be rather
typical situation in astrophysics, for instance in neutron stars,
the influence of photon dispersion in plasma on the $SL\nu$
process can be neglected and the threshold generated by the
non-zero plasmon mass should not be taken into account. However,
the method of exact solutions of Modified Dirac equation provides
us with analytical expressions for probability and intensity in
the whole range of possible parameters.

\acknowledgments One of the authors (A.S.) is thankful to Giorgio
Bellettini, Giorgio Chiarelli, Mario Greco and Gino Isidori for
the invitation to participate in Les Rencontres de Physique de la
Vallee d'Aoste on Results and Perspectives in Particle Physics.

This work has been supported by RFBR grant 11-02-01509-a.

\end{document}